\documentclass[prl,aps,preprint,amsmath,showpacs]{revtex4}

\usepackage{graphicx}% Include figure files
\usepackage{dcolumn}% Align table columns on decimal point
\usepackage{bm}% bold math
\begin{document}

%\preprint{APS/123-QED}

\title{Temperature-dependent transformation thermotics: From switchable thermal cloaks to macroscopic thermal diodes}

%\title{Nonlinear transformation thermotics: Macro thermal switch and diode}

%\title{Macro thermal diode with structural symmetry???}

%\title{Nonlinear transformation thermotics: Thermal switching and rectification on a macro scale}

%\title{Macro thermal diode using transformation media}

%\title{Macro thermal diode based on a thermal cloak with switching phenomena}

%\title{Thermal Computer: Starting from Nonlinear Thermal Metamaterials with Switching Phenomena}
%\title{Nonlinear Thermal Metamaterials: Switching Phenomena}% Force line breaks with \\
%\thanks{}%

%\author{Ying Li$^{1,2}$, Xiangying Shen$^1$, Yushan Ni$^2$, and Jiping Huang$^1$}\email{jphuang@fudan.edu.cn}

%\affiliation

%\address{$^1$Department of Physics, State Key Laboratory of Surface Physics, and Key Laboratory of Micro and Nano Photonic Structures (Ministry of Education), Fudan University, Shanghai 200433, China\\
% $^2$Department of Mechanics and Engineering Science, Fudan University, Shanghai 200433, China
% }

\author{Ying Li$^{2}$}
\author{Xiangying Shen$^{1}$}
\author{Zuhui Wu$^{1}$}
\author{Junying Huang$^{1}$}
\author{Yixuan Chen$^{1}$}
\author{Yushan Ni$^{2}$}
\author{Jiping Huang$^{1,}$}\email{jphuang@fudan.edu.cn.}

\affiliation{$^1$Department of Physics, State Key Laboratory of Surface Physics, Key Laboratory of Micro and Nano Photonic Structures (Ministry of Education), and Collaborative Innovation Center of Advanced Microstructures, Fudan University, Shanghai 200433, China\\
$^2$Department of Mechanics and Engineering Science, Fudan University, Shanghai 200433, China}

%\author{Ying Li}
%\affiliation{%
% Department of Physics, State Key Laboratory of Surface Physics, and Key Laboratory of Micro and Nano Photonic Structures (Ministry of Education), Fudan University, Shanghai 200433, China}%
%\affiliation{Department of Mechanics and Engineering Science, Fudan University, Shanghai 200433, China}%Lines break automatically or can be forced with \\
%\author{X. Y. Shen}%
 %\footnote{The first two authors have the same contribution.}
% \affiliation{%
% Department of Physics, State Key Laboratory of Surface Physics, and Key Laboratory of Micro and Nano Photonic Structures (Ministry of Education), Fudan University, Shanghai 200433, China}%

%\author{Yushan Ni}
% \email{niyushan@fudan.edu.cn}
%\affiliation{Department of Mechanics and Engineering Science, Fudan University, Shanghai 200433, China}

%\author{J. P. Huang}
% \email{jphuang@fudan.edu.cn; Tel.: +86 21 55665227; fax: +86 21 55665239.}
%\affiliation{%
% Department of Physics, State Key Laboratory of Surface Physics, and Key Laboratory of Micro and Nano Photonic Structures (Ministry of Education), Fudan University, Shanghai 200433, China}%

\date{\today}% It is always \today, today,

\begin{abstract}

The macroscopic control of ubiquitous heat flow remains poorly explored due to the lack of a fundamental theoretical method. Here, by establishing temperature-dependent transformation thermotics  for treating materials whose conductivity depends on temperature, we show analytical and simulation evidence for switchable thermal cloaking and a macroscopic thermal diode based on the cloaking. The latter allows heat flow in one direction but prohibits the flow in the opposite direction, which is also confirmed by our experiments. Our results suggest that the temperature-dependent transformation thermotics could be a fundamental theoretical method for achieving macroscopic heat rectification, and provide guidance both for macroscopic control of heat flow and for the design of the counterparts of switchable thermal cloaks or macroscopic thermal diodes in other fields like seismology, acoustics, electromagnetics, or matter waves.

%preservation or dissipation
%other fields, say, in   novel properties, not only in thermotics but also in other fields ranging from electromagnetic to acoustics.

\end{abstract}


%\pacs{81.05.Xj, 74.25.fc, 44.90.+c}% PACS, the Physics and Astronomy
%\keywords{Transformation thermotics; Thermal metamaterial; Thermal conductivity}%Use showkeys class option if keyword

\maketitle

%\tableofcontents

%Since all forms of energy can be completely converted to heat,

Heat flow is a ubiquitous phenomenon in nature, and hence how to control the flow of heat at will is of particular importance for human life. Fortunately, the past years have witnessed the possibility of manipulating
phononic~\cite{LiPRL04,ChangS06,WangPRL07,WangPRL08,LiRMP12,MaldovanN13,ChenNC14} or electronic \cite{GiazottoNat12,MartinezNC14,FornieriAPL14,MartinezNN15} heat conduction {\it at the
nanoscale}, which provides a promising method to make use of heat flux. So far, significant
progresses have been made, such as computation \cite{WangPRL07}, information storage
\cite{WangPRL08} and caloritronics \cite{GiazottoNat12,MartinezNC14}.
%The rectification of
%heat conduction at the macroscale is also proposed
%\cite{KobayashiAPL09,GoJHT10,KobayashiAPX12}.
%
%still far from being satisfactory,
%which prohibits macro thermal rectification
%
%for diverse thermal management problems, e.g., efficient refrigerators, solar
%cells, and energy-saving buildings. A reason behind the situation is due to
%the lack of a fundamental theoretical method.
On the contrary, steering heat conduction
{\it at the macroscale} is still far from being satisfactory~\cite{WangNanoLett14}, which prohibits macroscopic thermal rectification for diverse thermal management problems, e.g., efficient refrigerators, solar cells, and energy-saving buildings~\cite{KobayashiAPL09,GoJHT10,KobayashiAPX12}. A reason behind the situation is due to the lack of a fundamental theoretical method.

In 2008, Fan {\it et al.}~\cite{fan2008shaped}  adopted a coordinate transformation approach to propose a class
of thermal metamaterial where heat is caused to flow around an ``invisible'' region  at steady state, thus called {\it thermal cloaking}.
The cloaking originates from the fact that the thermal conduction equation remains form-invariant
under coordinate transformation. So far, the theoretical proposal of steady-state thermal cloaking~\cite{fan2008shaped} and its extensions (say, bifunctional cloaking of heat and electricity~\cite{LiJAP10} or nonsteady-state thermal cloaking~\cite{GuenneauOE12}) have been experimentally verified and developed~\cite{Narprl2012,SchittnyPRL13,XuPRL14,HanPRL14,MaPRL14}. The theoretical treatment based on coordinate transformation~\cite{fan2008shaped,LiJAP10,GuenneauOE12,LeonhardtS06,PendryS06,ChenNM10,nm2013,KadicRPP13},
which is called {\it transformation thermotics (or transformation thermodynamics)}, has a potential to become
a fundamental theoretical method for macroscopically manipulating heat flow at will.
%is  a powerful tool to manipulate  heat flow at will.
%ux and temperature signature.
%However, the existing
% transformation thermotics works for materials whose conductivity is independent of temperature (thus called {\it linear materials}).
%
% a  thermal counterpart of transformation optics~\cite{leonhardtsci2006,pendrysci2006,ChenNM10}, which is
%

However, in order to realize desired macroscopic thermal rectification, say, switchable thermal cloaking and macroscopic
 thermal diodes, the existing transformation thermotics~\cite{fan2008shaped,LiJAP10,GuenneauOE12} is not enough since it only holds for materials whose conductivity is independent of temperature (thus called {\it linear materials}).
For instance, the desired thermal diode should conduct heat in one direction, but insulate the heat in the opposite direction. This is a kind of asymmetric behavior of heat
current. For this purpose, nonlinear materials (whose conductivity relies on temperature) must be adopted. Actually, it is long known that for many materials, their thermal conductivities ($\kappa$) vary with temperature ($T$):
%\begin{itemize}
 (1) {\it $\kappa$ increases as $T$ increases.} Say, for noncrystalline solids, a series of experiments on glass \cite{PhysRevB.4.2029} showed that at low temperature the thermal conductivity is proportional to $T^{1.6}\sim T^{1.8}$;
(2) {\it $\kappa$ increases as $T$ decreases.} For example, measurements on single crystals of silicon
and germanium from $3\,\mathrm{K}$ to their melting point \cite{GlassbrennerPR63} showed that their thermal
conductivity decreases faster than $1/T$.
%\end{itemize}

{\bf Temperature-dependent transformation thermotics}

Now we are in a position to establish a theory of transformation thermotics for treating nonlinear materials, thus called {\it temperature-dependent transformation thermotics}. The details of the theory are given in Part I of Ref.~\cite{SUM},
which yield the key formula, Eq.~(S8).
%Here, we simply start from Eq.~(S8) in the Appendix.
 %the following equation, which is a part of  Eq.~(\ref{eq21}) in the Appendix,
%\begin{equation}
%  \tilde{\kappa}(T)=\frac{\tilde{J}(T)\kappa_0\tilde{J}^{\mathrm{t}}(T)}{\mathrm{det}[\tilde{J}(T)]}.
%  \label{eq:main}
%\end{equation}
This Eq.~(S8) denotes that instead of constructing materials for a background whose thermal conductivity is $T$-dependent, we may apply a $T$-involved transformation to the original thermal conduction equation. This process allows us to design switchable thermal cloaks and then macroscopic thermal diodes. The former serve as an extension of the extensively investigated cloaks without switches~\cite{fan2008shaped,LiJAP10,GuenneauOE12,Narprl2012,SchittnyPRL13,HanSR13,XuPRL14,HanPRL14,MaPRL14}; the latter are actually a useful application of the former in this work. We proceed as follows.

{\bf Switchable thermal cloaks}

%A natural application of a $T$-involved transformation is to add a working range for the device, which means that the device can be activated and deactivated at different temperatures. Here we manage to realize such a function for a typical
%thermal metamaterial, the thermal cloak.

{\it Design.}---
A traditional thermal cloak can protect a central region from an external heat flux and permits the region to remain a constant temperature  without disturbing the temperature distribution outside the cloak (Fig.~1). To achieve this cloaking effect, a simple radial stretch transformation of polar coordinates may be performed. As schematically shown in Fig.~1, the circular region with radius $R_2$ is compressed to the annulus region with radius in-between  $R_1$ and $R_2$, and the geometrical transformation can be written as
\begin{equation}
r'=r\frac{R_{2}-R_{1}}{R_{2}}+R_1\label{trans1},
\end{equation}
where $r\in[0,R_2]$ and $r'\in[R_1,R_2]$. Here $r'$ is the radial coordinate in physical space.

In order to realize different responses to heat flow on the two sides of a thermal diode, we need two types of thermal cloaks: one functions  at high
temperature (hereafter indicated as type-A cloaks), the other works  at low temperature (type-B cloaks). Unlike some previous proposed switchable electromagnetic cloaks
\cite{ChenACS11,ZhangIEEE13,WangPRB14}, the switching effect
should be triggered automatically by temperature changes.
For this purpose, an idea is to
modify  Eq.~(\ref{trans1}) as
\begin{equation}
   r'=r\frac{R_{2}-\tilde{R}_{1}(T)}{R_{2}}+\tilde{R}_{1}(T),\label{trans3}
\end{equation}
where $\tilde{R}_{1}(T)=R_1[1-(1+\mathrm{e}^{\beta(T-T_c)})^{-1}]$ for type-A cloaks and
$\tilde{R}_{1}(T)=R_1/(1+\mathrm{e}^{\beta(T-T_c)})$ for type-B cloaks. Here $T_c$ is a critical temperature around
which the cloak is switched on or off, and $\beta$ is a scaling coefficient which is set to be $2.5\,\mathrm{K}^{-1}$ in this work.
%
%The original function $r'$ maps $r\in[0,R_2]$ to $r'\in[R_1,R_2]$.
%For type-A cloaks, now we need to find a function $r'$ which still holds the original mapping at high temperature but becomes an identical mapping at low temperature. The following transformation meets the desired property,
%\begin{equation}
%   r'=r\frac{R_{2}-\tilde{R}_{1}(T)}{R_{2}}+\tilde{R}_{1}(T),\label{trans3}
%\end{equation}
%where $\tilde{R}_{1}(T)=R_1[1-(1+\mathrm{e}^{\beta(T-T_c)})^{-1}]$. Similarly,
%the $\tilde{R}_{1}(T)$ for type-B cloaks is $R_1/(1+\mathrm{e}^{\beta(T-T_c)})$.

So far, for obtaining thermal cloaks with switching phenomena, we need to combine Eq.~(\ref{trans3}) and Eq.~(S8). As a result, for the area with radius $r'\in[R_1,R_2]$  in Fig.~1, we achieve the thermal conductivities in polar coordinates, diag[$\tilde{\kappa}_r(T)$,$\tilde{\kappa}_{\theta}(T)$], as
\begin{equation}
   \tilde{\kappa}_r(T)=\kappa_0\dfrac{r'-\tilde{R}_1(T)}{r'},\ \tilde{\kappa}_{\theta}(T)=\kappa_0\dfrac{r'}{r'-\tilde{R}_1(T)}.
   \label{eq:04}
\end{equation}
Here $\kappa_0$ represents the $T$-independent thermal conductivity of the background.

{\it Finite-element simulations.}---
Then we perform finite element simulations based on the commercial software COMSOL Multiphysics. A thermal cloak with $R_1=1\,\mathrm{cm}$ and $R_2=2\,\mathrm{cm}$ is set in a box with size $8\,\mathrm{cm}\times 7\,\mathrm{cm}$ as shown in Fig.~2. Heat diffuses from the left boundary with high temperature $T_H$ to the right boundary with low temperature $T_L$. Meanwhile, the upper and lower boundaries of the simulation box are thermally isolated.

%Without loss of generality,  we choose  $L_x/8$ ($L_x$: width of the simulation box),  the initial temperature and the background's thermal conductivity  as the units of length, temperature and thermal conductivity, respectively. In doing so, the scales of length,  temperature and thermal conductivity are nondimensionalized accordingly.
%The unit of energy is defined by multiplying unit temperature with $1\,\mathrm{J/K}$.

The simulation results of a type-A cloak are shown in Fig.~2(a,b). In Fig.~2(a), at high temperature ($T=340\,\mathrm{K}\sim 360\,\mathrm{K}$), we observe that the cloak is functioning and thermally hiding
the object located at the central region with radius $R_1$.
However, when the environment is changed to  low temperature ($T=300\,\mathrm{K}\sim 320\,\mathrm{K}$), the cloak is turned off.
That is, the temperature distribution outside the object is distorted, just as the cloak (located between $R_1$ and $R_2$)
is absent. Owing to the antisymmetry between type-A and type-B cloaks,
the type-B cloaks exhibit the behavior similar to Fig.~2(a,b),
but switching on (or off) at low (or high) temperature.
 % and switching off at high temperature.
%The $\kappa_r^{\mathrm{Sw}}$ and $\kappa_{\theta}^{\mathrm{Sw}}$ in Eq.~(\ref{eq:4}) with $T$ and $r'$ as independent variables are presented in Fig.~3.

% The
%critical temperature $T_c$ is set to be $1.5$.

{\it Theoretical realization based on the effective medium theory.}---
The materials designed according to Eq.~(\ref{eq:04}) is anisotropic
and inhomogeneous, which is difficult to be realized in experiments.
In fact, for constructing such materials, we can simply utilize alternating
layers of two homogeneous isotropic sub-layers with thicknesses $d_1$ and
$d_2$ and conductivities $\kappa_a$ and $\kappa_b$. For simplification, in
this work, we set $d_1=d_2$. According to the theoretical analysis and
effective medium theory
(EMT)~\cite{Narprl2012,GuenneauOE12,LiJAP10,Huangpr06,HanSR13}, the
conductivities of two sub-layers should satisfy
$\kappa_a\kappa_b\approx\kappa_0^2$ for a traditional cloak.
In order to endue  conventional cloaks with the switching effect, some mathematical operations must be carried out on $\kappa_a$ and $\kappa_b$ in the way analogous to what we did on $\tilde{R}_1(T)$.
Therefore, we obtain $\kappa_1(T)$ and $\kappa_2(T)$ as the new
temperature-dependent thermal conductivities of the sub-layers,
\begin{equation}
%\begin{array}{lll}
\kappa_1(T)=\kappa_a+\dfrac{\kappa_0-\kappa_a}{1+\mathrm{e}^{\beta(T-T_C)}},
\kappa_2(T)=\kappa_b-\dfrac{\kappa_b-\kappa_0}{1+\mathrm{e}^{\beta(T-T_C)}}.
\label{eq:05}
%\end{array}
\end{equation}
The two expressions yield:  $\kappa_1(T)\rightarrow\kappa_a$ and $\kappa_2(T)\rightarrow\kappa_b$ when $T\gg T_C$;  $\kappa_1(T)\rightarrow \kappa_0$ and $\kappa_2(T)\rightarrow \kappa_0$ as $T\ll T_C$. (The detailed dependence of $\kappa_1(T)$ and $\kappa_2(T)$ on temperature is displayed in Fig.~S3 of Part III of Ref.~\cite{SUM}.) That is,  the cloaking effect is switched on (or off) for high (or low) temperature environment $T\gg T_C$ (or $T\ll T_C$). Eq.~(\ref{eq:05}) offers a convenient tool to help experimentally realize our theoretical design of Eq.~(\ref{eq:04}).  Next, we
plot Fig.~2(c,d), which shows the simulation results of 10 alternating layers
for constructing type-A cloaks. Evidently, Fig.~2(c,d) displays the
switching phenomenon similar to Fig.~2(a,b). The procedure holds the same for achieving
type-B cloaks.

{\bf Macro thermal diode}

{\it Design.}---
The above  thermal cloaking may help to design a kind of
macroscopic thermal diodes. As shown in Fig.~3(a), the diode device contains
Regions I, II and III: Region I (II) is a segment of the type-A (type-B) cloak, and Region III is a thermal conductor. Compared
with a full system, the cloaking effect still exists in our diode but is not
perfect. There will be a small amount of heat flux conducting through the
central region for the insulating case. However, the truncation of a whole
cloak is necessary to separate the type-A part and the type-B part. The
antisymmetry of type-A and type-B cloaks is expected to cause different
behaviors of heat conducting from the two opposite directions. The transformation
plays an important role in introducing anisotropic effect to the structure,
which guides the heat flux to the boundaries to enhance the thermal insulating
effect.

%Region I to Region II and vice versa.

%, when the environment temperature is around $T_c$.

% Cloak segments with inner radius $3.6$
%and outer radius $4$ are fixed in a simulation box of size $8\times 4$, with the upper and lower boundaries
%thermal insulated. The thermal conductivities of the background and the central thermal conductor are
%$1$ and $10$, respectively. Temperatures of the left and right boundaries
%($T_l$ and $T_r$) are set to have an average $T_c$.

{\it Finite-element simulations.}---
Then we perform finite element simulations.  Fig.~3(c,d) shows the simulation results of the device, which helps to
insulate heat  from right to left but conduct the heat from left to right. That is, the behavior of diode can be achieved indeed due to the antisymmetry of type-A and type-B cloaks (namely, Region I and Region II).
%For the convenience of future experimental demonstration, here we also perform simulations based on the EMT [Eq.~(\ref{eq:05})] with 10 alternating layers, each having two homogeneous isotropic sub-layers. The same effect of diode is achieved; see Fig.~4(c) and (d).
Moreover, for different temperature biases (obtained by subtracting the temperature at the left boundary from that at the right boundary),
$\Delta T$, we also calculate the total heat current $J$ by integrating the $x$ component of heat flux across the line $x = 0$; see Fig.~3(b). The device
displays a significant rectifying effect, which has a maximum rectification ratio of $30$ for the current parameter set.

% of Fig.~3(b).

{\it Experimental realization based on the effective medium theory.}---
In order to realize such a  macroscopic thermal diode, we can also adopt
the EMT. As discussed above, the switchable thermal cloak can be constructed
with alternating layers. The thermal conductivities of the sub-layers are
required to be sensitive to the temperature change around the critical point.
This kind of behaviors can be found in the phase transitions of some materials
\cite{OhAPL10,ZhengNatComm11,SiegertRPP15}. However, inspired by the spirit of
metamaterials (yielding novel functions by assembling conventional materials into
specific structures), we manage to realize the macroscopic thermal diode with
materials of constant conductivities. Our method is that instead of directly
resorting to the transitions of materials' physical properties, we use the
structural transition to trigger the switching effect. According to the
demands of our design, the geometrical configuration of the device should
change rapidly as temperature varies. Fortunately,  we
found that the shape-memory alloy (SMA)~\cite{OmoriNat13} may be able to help. As shown in the schematic diagrams of our design [see Fig.~4(a) and (b);
more details of our experiment can be found in Part II of Ref.~\cite{SUM}], the sub-layers of the cloak
segments are coppers and expanded-polystyrene (EPS). Around the critical temperature, the
deformations of the SMA slices drive the copper slices to connect or
disconnect the copper layers. The connection and disconnection can be equivalently regarded
as transitions of the local thermal conductivities.  Thus a temperature-dependent thermal metamaterial is realized
with materials of constant thermal conductivities (a whole switchable thermal
cloak can also be built with the same method). The experimental results are
shown in Fig.~4(c) and (d). Fig.~4(c) displays the temperature distribution
within the central region, which is almost constant. That is, in this case, heat almost cannot  flow through this central region. Thus, Fig.~4(c) corresponds to the insulating case of the diode. In contrast, Fig.~4(d) represents the conducting case of the diode. This is because Fig.~4(d) shows a significant
temperature gradient within the central region. Also, the
temperature distribution appears to be horizontal. As a result, we can conclude that an evident heat flow comes to appear within the central region of Fig.~4(d).

%The strong contraction between
%Fig.~4(c) and (d) implies a good performance of the fabricated macro thermal
%diode.

%Meanwhile, the anisotropy
%of the layered structure is also broken or restored by the connection or
%disconnection.

{\bf Conclusions}

%In a word, we propose a more general transformation mapping theory for thermal conduction, where the parameters of materials vary with the temperature.
%A universal expression of transformed conductivity is given by using tensor analysis method. As the original transformation theory, it permits people to construct metamaterials with new properties by tailoring anisotropic and inhomogeneous materials appropriately. According to the nonlinear responses of materials, we also can design novel metamaterials based on the more general transformation mapping theory. As shown in the article, we give an idea to realize the switchable thermal metamaterials, which are prevented from functioning until the temperature reaches a certain value. Furthermore, a method to construct thermal diode with such switchable device is also put forward as a demonstration of its element role in future design  We believe our work implies the applications of transformation mapping theory on designing metamaterials with much larger variety in many domains of physics.

%It is a challenge to achieve macro thermal rectification for diverse thermal management problems, say, efficient refrigerators, solar cells, and energy-saving buildings. In the literature, transformation thermotics (or thermodynamics) has been extensively adopted to handle linear thermal metamaterials whose conductivity does not depend on temperature.

We have established a theory of temperature-dependent transformation thermotics  for dealing with  thermal materials whose conductivity is temperature-dependent. The theory serves as a fundamental theoretical method for designing switchable thermal cloaking. We have also shown that the switchable thermal cloaks can be employed for achieving a macroscopic thermal diode, which has also been experimentally realized by assembling homogeneous and isotropic materials according to the design based on the EMT (effective medium theory). The diode has plenty of potential applications related to heat preservation, heat dissipation, or even heat illusion~\cite{HanAM14,ZhuAIPA15} in many areas like efficient refrigerators, solar cells, energy-saving buildings, and military camouflage. Thus, by using temperature-dependent
transformation thermotics to tailor nonlinear effects appropriately, it becomes possible to achieve desired thermal metamaterials with diverse capacities for  macroscopic thermal rectification. On the same footing, our consideration (for cloaks and diodes) adopted in this work can be extended to obtain the counterparts of both switchable thermal cloaks and macroscopic thermal diodes in other fields like seismology, acoustics, electromagnetics, or matter waves.

%Besides finite element simulations, for experimental demonstration, we have also realized these materials (switch and diode) by using real materials according to the design based on the effective medium theory. Thus, by using nonlinear transformation thermotics to tailor nonlinear effects appropriately, it becomes possible to achieve desired thermal metamaterials with diverse kinds of rectifications. On the same footing, our consideration (for thermal switch and diode) as discussed in this work can be extended to obtain the counterparts of thermal switches and diodes in other fields like electromagnetic waves, acoustics waves, or matter waves.

%As a result of the  nonlinear transformation thermotics, we have achieved a new type of switchable thermal cloaks and hence macroscopic thermal rectification, namely, a macro thermal diode. The latter has plenty of potential applications related to heat preservation, heat dissipation, or even heat illusion~\cite{HanAM14,ZhuAIPA15}.

%To make our work more general, reduced parameters are used for the convenience of future experimental demonstration: namely, all thermal conductivities have a unit of that of the background, thus $\kappa_0 =1$; length parameters have a unit of $R_1$, thus $R_1=1$.

{\it Acknowledgement.} We acknowledge the financial support by the National Natural Science Foundation of China under Grants No.~11222544 and No. 1157209 (Y.L. and Y.N.), by the Fok Ying Tung Education Foundation under Grant No. 131008, by the Program for New Century Excellent Talents in University (NCET-12-0121), and by the CNKBRSF under Grant No. 2011CB922004.

Y.L. and X.S. contributed equally to this work.

\clearpage
\newpage
%Figure Captions

Fig.~1 (color online). Schematic graph depicting a thermal cloak between radius $R_1$ and  $R_2$. Red lines with arrows denote the flow of heat: the cloak  does not disturb the heat flow at the region with radius larger than $R_2$; the heat flux cannot enter the central region with radius smaller than $R_1$.

%between $R_1$ and $R_2$

%Fig.~2. Heat flows from left to right on a cloak with $R_1=0.01$\,m and $R_2=0.02$\,m in Wiedemann--Franz model. The temperature on the left is $200$\,K and the right is $100$\,K. Snapshots of temperature distribution: (a) $t=1$\,s, (b)$t=5$\,s, (c)$t=10$\,s, (d)$t=15$\,s.

%Fig.~3. Heat diffuses from left to right on a concentrator with $R_1=0.01$\,m, $R_2=0.02$\,m, and $R_3=0.005$\,m in Wiedemann--Franz model. The temperature on the left is $200$\,K and the right is $100$\,K. Snapshots of temperature distribution: (a) $t=0.5$\,s, (b)$t=1$\,s, (c)$t=2$\,s, (d)$t=5$\,s.

%Fig.~4.  Diffusion of heat from left to right on a cloak with $R_1=0.01$\,m and $R_2=0.02$\,m in Callaway's model. The temperature on the left is $50$\,K and the right is $20$\,K. Snapshots of temperature distribution: (a) $t=25$\,s, (b)$t=50$\,s, (c)$t=75$\,s, (d)$t=100$\,s.

%Fig.~5. Conduction of heat from left to right on a concentrator with $R_1=0.01$\,m, $R_2=0.02$\,m, and $R_3=0.005$\,m in Callaway's model. The temperature on the left is $50$\,K and the right is $20$\,K. Snapshots of temperature distribution: (a) $t=25$\,s, (b)$t=50$\,s, (c)$t=75$\,s, (d)$t=100$\,s.

%Fig.~7. $x$-dependent temperature ($T$) along $y=0$. (a) gives the temperature data of thermal metamaterials and background in Wiedemann--Franz law's case. (b) presents the temperature data of thermal metamaterials and background in Callaway's model.

Fig.~2 (color online). Switchable thermal cloaks obtained by two-dimensional finite-element simulations: (a,c) switch on for the  temperature above $340\,\mathrm{K}$ and (b,d) switch off for the  temperature below $320\,\mathrm{K}$. The color surface denotes the distribution of temperature, where isothermal lines are indicated; heat diffuses from left to right; the upper and lower boundaries are thermal insulation. (a) and (b) show the results of thermal conductivities calculated according to Eq.~(\ref{eq:04}); (c) and (d) show the results of 10 alternating layers of two sub-layers with $\kappa_1(T)$ and $\kappa_2(T)$ given by Eq.~(\ref{eq:05}) (EMT).  In (a-d), an object with thermal conductivity $0.01\,\mathrm{W/mK}$ is set in the central region with radius $R_1$. Parameters:  $\kappa_0 = 1\,\mathrm{W/mK}$, $\kappa_a=0.1\,\mathrm{W/mK}$, $\kappa_b=10\,\mathrm{W/mK}$, $R_1 =1\,\mathrm{cm}$, $R_2=2\,\mathrm{cm}$, and $T_c = 330\,\mathrm{K}$.

%(c) the thermal concentrator is ``closed'' when temperature is below the $140$ K. (d) the thermal concentrator is ``open'' when temperature is above the $160$ K.

%Fig.~3 The components of thermal conductivities of switchable thermal cloak are shown. (a) The $\kappa^{\mathrm{Sw}}_{r}$ of the switchable thermal cloak, (b) the $\kappa^{\mathrm{Sw}}_{\theta}$ of the switchable thermal cloak.

Fig.~3 (color online). (a) Sketch of a thermal diode, which is the rectangular area
enclosed by the solid black lines. The blurred area outside is a reference
and actually does not exist in the design. I, II, and III represent three regions, respectively. Here the arrows indicate the direction of
heat flow; the length of arrows represents the amount of heat flux: the heat
flux transferred from right to left (upper panel: the insulating case) is much
smaller than that from left to right (lower panel: the conducting case).  (b)
Heat current $J$ versus temperature bias $\Delta T$. (c,d) Thermal diode
obtained by two-dimensional finite-element simulations: (c) the insulating case
and (d) the conducting case. The color surface denotes the distribution of
temperature; white arrows represent the direction of heat flow; the length of
white arrows indicates the amount of heat flux; the upper and lower boundaries
are thermal insulation. Thermal conductivities are calculated according to
Eq.~(\ref{eq:04}); an object with thermal conductivity $10\,\mathrm{W/mK}$ is
set in the central region with radius  $R_1$. Parameters:  $\kappa_0 =
1\,\mathrm{W/mK}$, $R_1 =3.6\,\mathrm{cm}$,  $R_2=4\,\mathrm{cm}$, and $T_c =
330\,\mathrm{K}$.

%(which is obtained by subtracting the temperature at the right boundary from that at the left boundary).

Fig.~4 (color online). (a,b) Scheme of  experimental demonstration of the macroscopic thermal
diode: (a)  insulating case and (b)  conducting case. Both the copper-made
concentric layered structure and the central copper plate (both displayed in
orange) are placed on an expanded-polystyrene (EPS) plate which is not
shown for clarity. The left and right sides of the diode are stuck in
water to have constant temperature boundary conditions. (a) When cold water
 is filled in the left container (light blue)
and hot water  the right container (pale red), the
bimetallic strips of SMA and copper (white) warp up and
the device blocks heat from right to left. (b) When the two containers swap their
locations, the bimetallic strips (white) flatten and the device conducts heat from left
to right. (c,d) Experimentally measured temperature distribution of the
device: (c)  insulating case and (d)  conducting case.

%Temperature distribution for (a) the insulated case when heat conducts from left to right; and (b) the conducting case when heat diffuses from right to left. The amount of currency is represented by the white arrows. (c) and (d) are simulation results by utilizing the EMT.

\clearpage
\newpage
\begin{figure*}[t]
\begin{center}
\centerline{\includegraphics[width=\linewidth]{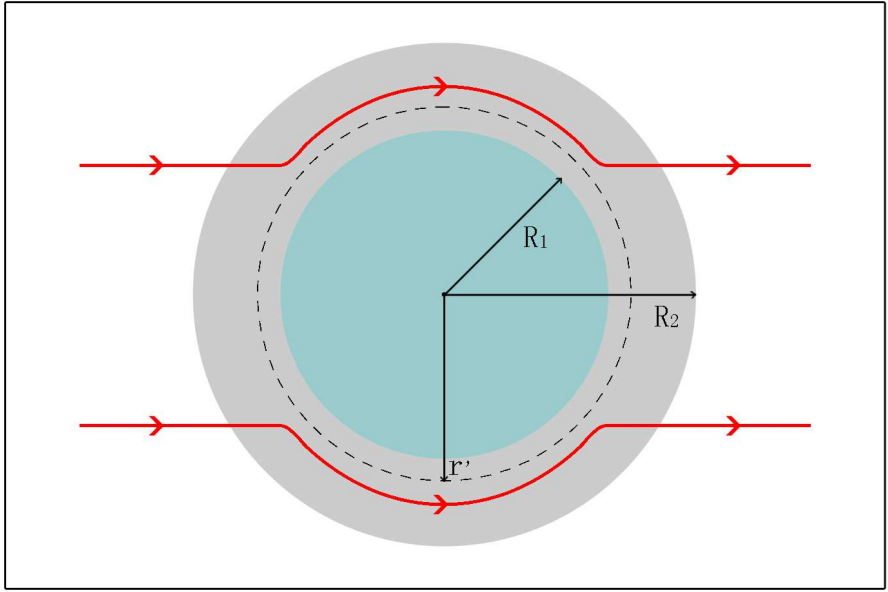}}
\caption{/Li, Shen, Wu, Huang, Chen, Ni, and Huang}
\end{center}
\end{figure*}

\clearpage
\newpage
\begin{figure*}[t]
\begin{center}
\centerline{\includegraphics[width=\linewidth]{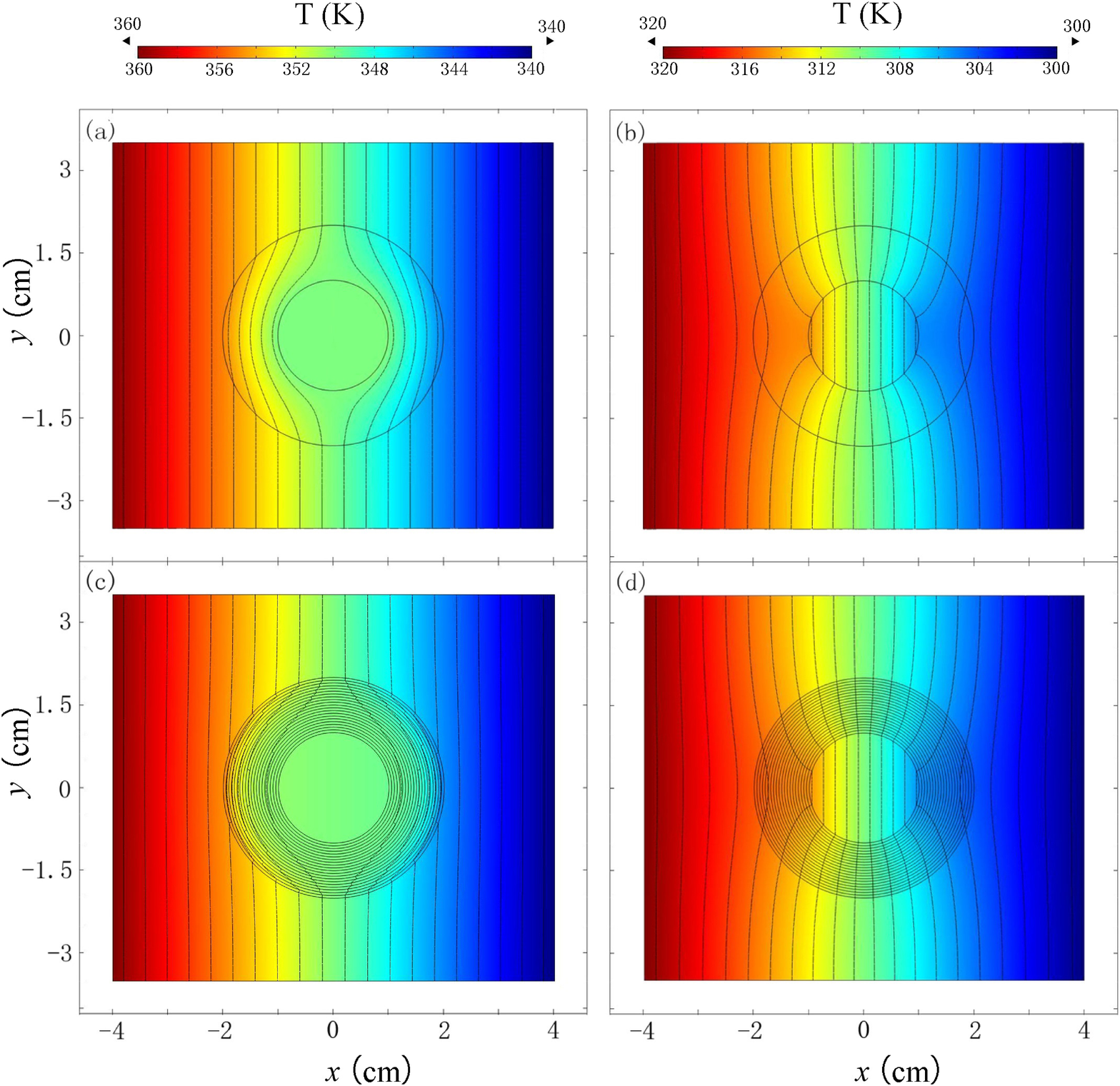}}
\caption{/Li, Shen, Wu, Huang, Chen, Ni, and Huang}
\end{center}
\end{figure*}

%\clearpage
%\newpage
%\begin{figure*}[t]
%\begin{center}
%\centerline{\includegraphics[width=\linewidth]{fig3.eps}}
%\caption{/Li, Shen, Ni, and Huang}
%\end{center}
%\end{figure*}

\clearpage
\newpage
\begin{figure*}[t]
\begin{center}
\centerline{\includegraphics[width=\linewidth]{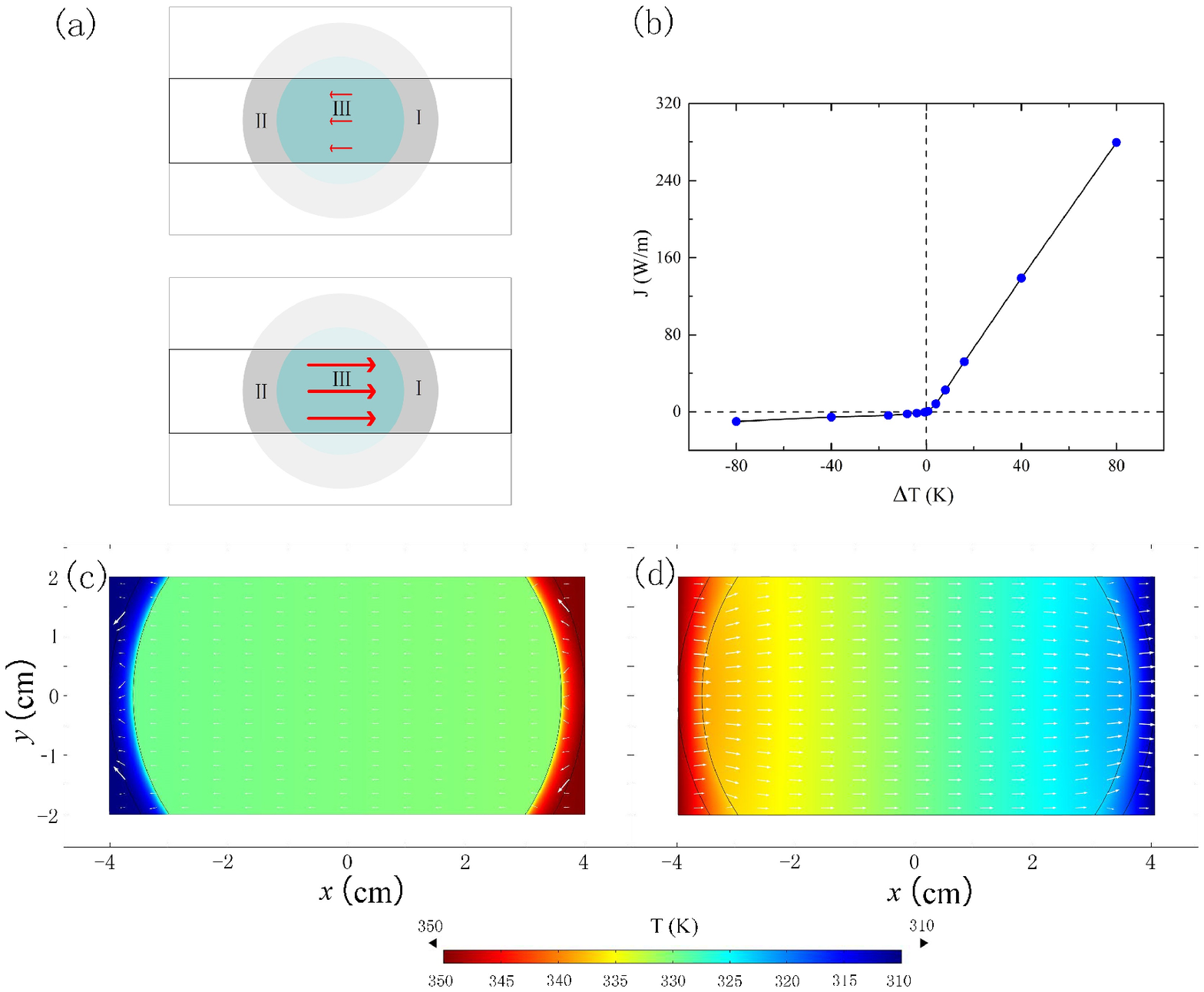}}
\caption{/Li, Shen, Wu, Huang, Chen, Ni, and Huang}
\end{center}
\end{figure*}

\clearpage
\newpage
\begin{figure*}[t]
\begin{center}
\centerline{\includegraphics[width=\linewidth]{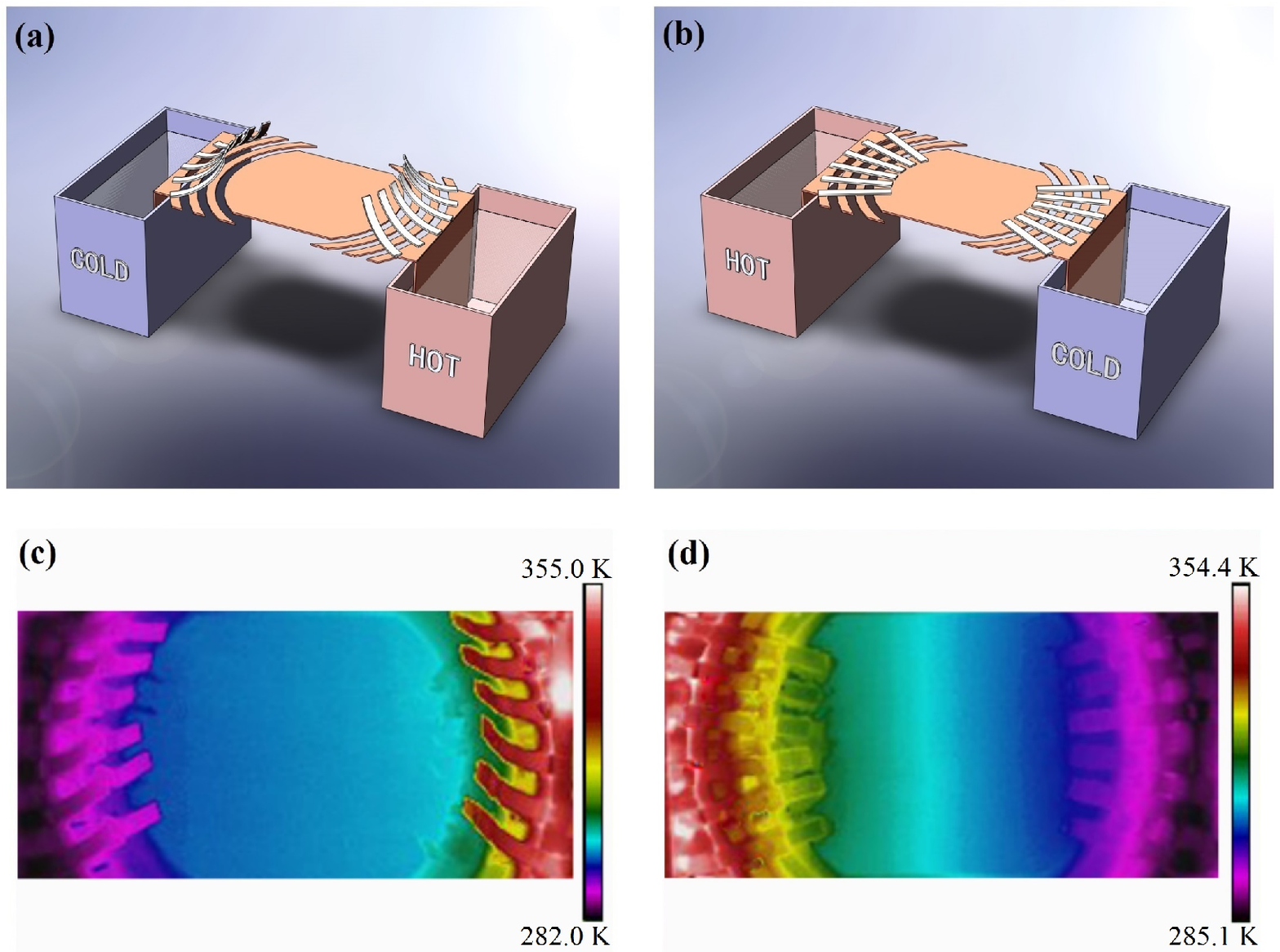}}
\caption{/Li, Shen, Wu, Huang, Chen, Ni, and Huang}
\end{center}
\end{figure*}


\begin{thebibliography}{99}
%\bibliography{}% Produces the bibliography via BibTeX.

\bibitem{LiPRL04} B. Li, L. Wang, and G. Casati, Phys. Rev. Lett. {\bf 93}, 184301 (2004).
 %Thermal Diode: Rectification of Heat Flux

\bibitem{ChangS06} C. W. Chang, D. Okawa, A. Majumdar, and A. Zettl, Science {\bf 314}, 1121 (2006).
    %Solid-State Thermal Rectifier

\bibitem{WangPRL07} L. Wang and B. Li,  Phys. Rev. Lett. {\bf 99}, 177208 (2007).
%Thermal Logic Gates: Computation with Phonons

\bibitem{WangPRL08} L. Wang and B. Li, Phys. Rev. Lett. {\bf 101}, 267203 (2008).
%â€œThermal Memory: A Storage of Phononic Information

\bibitem{LiRMP12} N. Li, J. Ren, L. Wang, G. Zhang, P. Hanggi, and B. Li, Rev. Mod. Phys. {\bf 84}, 1045 (2012).
% Phononics: Manipulating heat flow with electronic analogs and beyond

\bibitem{MaldovanN13} M. Maldovan, Nature {\bf 503}, 209 (2013).
%Sound and heat revolutions in phononics

\bibitem{ChenNC14} Z. Chen, C. Wong, S. Lubner, S. Yee, J. Miller, W. Jang, C. Hardin, A. Fong, J. E. Garay, and C. Dames, Nature Communications {\bf 5}, 5446 (2014).
%A photon thermal diode

\bibitem{GiazottoNat12} F. Giazotto and  M. J. Mart{\' i}nez-P{\' e}rez, Nature {\bf 492}, 401 (2012).
%The Josephson heat interferometer

\bibitem{MartinezNC14} M. J. Mart{\'{i}}nez-P{\'{e}}rez and F. Giazotto,  Nature Communications {\bf 5}, 3579 (2014).
%A quantum diffractor for thermal flux

\bibitem{FornieriAPL14} A. Fornieri, M. J. Mart{\' i}nez-P{\' e}rez, and F. Giazotto,  Appl. Phys. Lett. {\bf 104}, 183108 (2014).
%A normal metal tunnel-junction heat diode

\bibitem{MartinezNN15} M. J. Mart{\' i}nez-P{\' e}rez, A. Fornieri, and F. Giazotto,  Nature Nanotech. {\bf 10}, 303 (2015).
%Rectification of electronic heat current by a hybrid thermal diode

\bibitem{WangNanoLett14} Y. Wang, A. Vallabhaneni, J. Hu, B. Qiu, Y. P. Chen, and X. Ruan, Nano Lett. {\bf 14}, 592 (2014).

\bibitem{KobayashiAPL09} W. Kobayashi, Y. Teraoka, and I. Terasaki, Appl. Phys. Lett. {\bf 95}, 171905 (2009).
%An oxide thermal rectifier

\bibitem{GoJHT10} D. Go and M. Sen, J. Heat Trans. {\bf 132}, 124502 (2010).
%On the Condition for Thermal Rectification Using Bulk Materials

\bibitem{KobayashiAPX12} W. Kobayashi, D. Sawaki, T. Omura, T. Katsufuji, Y. Moritomo, and I. Terasaki, Appl. Phys. Express {\bf 5}, 027302 (2012).
%Thermal Rectification in the Vicinity of a Structural Phase Transition

\bibitem{fan2008shaped}
C.~Z. Fan, Y.~Gao, and J.~P. Huang, Appl. Phys. Lett. \textbf{92}, 251907 (2008).

\bibitem{LiJAP10} J. Y. Li, Y. Gao, and J. P. Huang, J. Appl. Phys. {\bf 108}, 074504 (2010).

\bibitem{GuenneauOE12} S. Guenneau, C. Amra, and D. Veynante, Optics Express {\bf 20}, 8207 (2012).

\bibitem{Narprl2012} S. Narayana and Y. Sato, Phys. Rev. Lett. \textbf{108}, 214303 (2012).

\bibitem{SchittnyPRL13} R. Schittny, M. Kadic, S. Guenneau, and M. Wegener, Phys. Rev. Lett. {\bf 110}, 195901 (2013).
\bibitem{XuPRL14} H. Xu, X. Shi, F. Gao, H. Sun, and B. Zhang, Phys. Rev. Lett. {\bf 112}, 054301 (2014).

\bibitem{HanPRL14} T. Han, X. Bai, D. Gao, J. T. L. Thong, B. Li, and C.-W. Qiu, Phys. Rev. Lett. {\bf 112}, 054302 (2014).

\bibitem{MaPRL14} Y. Ma, Y. Liu, M. Raza, Y. Wang, and S. He, Phys. Rev. Lett. {\bf 113}, 205501 (2014).
\bibitem{LeonhardtS06} U. Leonhardt, Science {\bf 312}, 1777 (2006).

\bibitem{PendryS06} J. B. Pendry, D. Schurig, and D. R. Smith, Science {\bf 312}, 1780 (2006).

\bibitem{ChenNM10} H. Chen, C. T. Chan, and P. Sheng, Nature Materials {\bf 9}, 387 (2010).

\bibitem{nm2013} N. Landy and D. R. Smith, Nature Materials {\bf 12}, 25 (2013).

\bibitem{KadicRPP13} M. Kadic, T. B\"{u}ckmann, R. Schittny, and Martin Wegener, Rep. Prog. Phys. {\bf 76}, 126501 (2013).
%Metamaterials beyond electromagnetism

\bibitem{PhysRevB.4.2029} R. C. Zeller and R. O. Pohl, Phys. Rev. B \textbf{4}, 2029 (1971).
%\bibitem{PhysRev.113.1046}
%J. Callaway, Phys. Rev. \textbf{113}, 1046 (1959).

\bibitem{GlassbrennerPR63} C. J. Glassbrenner and G. A. Slack, Phys. Rev. {\bf 134}, A1058 (1964).
\bibitem{HanSR13} T. Han, T. Yuan, B. Li, and C. Qiu, Scientific Reports {\bf 3}, 1593 (2013).


\bibitem{ChenACS11} P. Y. Chen and A. Al{\` u}, ACS Nano {\bf 5}, 5855 (2011).

\bibitem{ZhangIEEE13} W. Zhang, W. M. Zhu, H. Cai, M. J. Tsai, G. Lo, D. P. Tsai, H. Tanoto, J. Teng, X. Zhang, D. Kwong, and A. Liu, IEEE J. Sel. Top. Quant. {\bf 19}, 4700306 (2013).

\bibitem{WangPRB14} R. F. Wang, Z. L. Mei, X. Y. Yang, X. Ma, and T. J. Cui, Phys. Rev. B {\bf 89}, 165108 (2014).




%\bibitem{li2010bifunctional}
%J.~Y. Li, Y.~Gao, and J.~P.~Huang, J. Appl. Phys. \textbf{108}, 074504 (2010).

\bibitem{Huangpr06} J. P. Huang and K. W. Yu, Phys. Rep. \textbf{431}, 87 (2006).
\bibitem{SUM} See Supplemental Material at http://link.aps.org/ for (Part I) temperature-dependent transformation thermotics, (Part II) our experimental demonstration of the macroscopic thermal diode,  and (Part III) the temperature dependence of  $\kappa_1$ and $\kappa_2$.

\bibitem{OhAPL10} D. Oh, C. Ko, S. Ramanathan, and D. G. Cahill, Appl. Phys. Lett. \textbf{96}, 151906 (2010).

\bibitem{ZhengNatComm11} R. Zheng, J. Gao, J. Wang, and G. Chen, Nat. Commun. \textbf{2}, 289 (2011).

\bibitem{SiegertRPP15} K. S. Siegert, F. R. L. Lange, E. R. Sittner, H. Volker, C. Schlockermann, T. Siegrist, and M. Wuttig, Rep. Prog. Phys. \textbf{78}, 013001 (2015).

\bibitem{OmoriNat13} T. Omori and R. Kainuma, Nature {\bf 502}, 7469 (2013).

\bibitem{HanAM14} T. C. Han, X. Bai, J. T. L. Thong, B. W. Li, and C. W. Qiu, Adv. Mat. {\bf 26}, 1731 (2014).

\bibitem{ZhuAIPA15} N. Q. Zhu, X. Y. Shen, and J. P. Huang, AIP Advances {\bf 5}, 053401 (2015).
% Converting the patterns of local heat flux via thermal illusion device


%\bibitem{leonhardtsci2006}
%U. Leonhardt, Science \textbf{312}, 1777 (2006).


%\bibitem{pendrysci2006}
%J. B. Pendry, D. Schurig, and D. R. Smith, Science \textbf{312}, 1780 (2006).



%\bibitem{schsci06}
%D. Schurig, J. J. Mock, B. J. Justice, S. A. Cummer, J. B. Pendry, A. F. Starr, and D. R. Smith, Science \textbf{314}, 977 (2006).

%\bibitem{Chenapl07}
%H. Y. Chen and C. T. Chan, Appl. Phys. Lett. \textbf{91}, 183518 (2007).

%\bibitem{Cummernjp07}
%S. A. Cummer and D. Schurig, New J. Phys. \textbf{9}, 45 (2007).

%\bibitem{Farprl08}
%M. Farhat, S. Enoch, S. Guenneau, and A. B. Movchan, Phys. Rev. Lett. \textbf{101}, 134501 (2008).

%\bibitem{Liuepjap09}
%B. Liu and J. P. Huang, Eur. Phys. J-Appl. Phys. \textbf{48}, 093901 (2009).

%\bibitem{Norprsl08}
%A. N. Norris, Proc. R. Soc. London, Ser. A: Mathematical, Physical and Engineering Science \textbf{464}, 2411 (2008).

%\bibitem{Zhaprl11}
%S. Zhang, C. G. Xia, and N. Fang, Phys. Rev. Lett. \textbf{106}, 024301 (2011).

%\bibitem{Bruapl09}
%M. Brun, S. Guenneau, and A. B. Movchan, Appl. Phys. Lett. \textbf{94}, 061903 (2009).

%\bibitem{Parapl12}
%W. J. Parnell, A. N. Norris, and T. Shearer, Appl. Phys. Lett. \textbf{100}, 171907 (2012).

%\bibitem{Farprl09}
%M. Farhat, S. Guenneau, and S. Enoch, Phys. Rev. Lett. \textbf{103}, 024301 (2009).

%\bibitem{Steprl12}
%N. Stenger, M. Wilhelm, and M. Wegener, Phys. Rev. Lett. \textbf{108}, 014301 (2012).

%\bibitem{Chenepl09}
%H. Y. Chen, J. Yang, J. Zi, and C. T. Chan, Europhys. Lett. \textbf{85}, 24004 (2009).

%\bibitem{Zhaprl08}
%S. Zhang, D. A. Genov, C. Sun, and X. Zhang, Phys. Rev. Lett. \textbf{100}, 123002 (2008).

%\bibitem{Greennjp08}
%A. Greenleaf, Y. Kurylev, M. Lassas, and G. Uhlmann, New J. Phys. \textbf{10}, 115024 (2008).

%\bibitem{Diajo11}
%A. Diatta and S. Guenneau, J. Opt. \textbf{13}, 024012 (2011).

%\bibitem{Shenijht2014} X. Y. Shen and J. P. Huang, Int. J. Heat Mass Transfer \textbf{78}, 1 (2014).

%\bibitem{Gaoepl2013} Y. Gao and J. P. Huang, EPL \textbf{104}, 44001 (2013).

%\bibitem{Gueoe2012} S. Guenneau, C. Amra, and D. Veynante, Opt. Express \textbf{20}, 8207 (2012).

\end{thebibliography}
\end{document}